\documentstyle [alife6,epsf] {article}

\newcommand{\boldsymbol}[1]{\mbox{\boldmath$#1$}}

\title{Large-scale evolution and extinction in a hierarchically
  structured environment }
\author{C. Wilke, S. Altmeyer\and T. Martinetz \\
Ruhr-Universit\"at Bochum\\
D-44780 Bochum, Germany\\
e-mail:  Claus.Wilke@neuroinformatik.ruhr-uni-bochum.de}

\begin{document}
\maketitle
\begin{abstract}
A class of models for large-scale evolution and mass extinctions is presented.
These models incorporate environmental changes on all scales, from influences
on a single species to global effects. This is a step towards a
unified picture of mass extinctions, which enables one to study
coevolutionary effects and 
external abiotic influences with the same means. The
generic features of such 
models are studied in a simple version, in which all environmental
changes are generated at random and without feedback from other parts
of the system. 
\end{abstract}

\section{Introduction}

In the history of the Earth, there have been several catastrophic
events which in a short period of time have wiped out large parts of
the existing species. The amount of species annihilated in such
events has been up to 96\%\ of the biodiversity at that
time~\cite{Raup86}. It has often been argued that these mass
extinctions must have been caused by 
some disastrous abiotic incidences like extraterrestrial impacts.
Evidence in favor of that has been put forward~\cite{Alvarez87}, but
on the other hand, 
only 5\%\ of the total loss of biodiversity in the fossil record can
be connected to mass extinctions. The rest are the so-called
background extinctions, which happen on much smaller
scales. Interestingly, the two types of extinction cannot clearly be
distinguished from another in the frequency distribution of extinction
event sizes. The event sizes' distribution forms a smooth curve,
very close to a power-law~\cite{Sole96}.

In order to explain a single smooth distribution, the idea of
coevolutionary avalanches has been developed~\cite{Kauffman92}. The
extinction of a 
single species might cause another species to die out, which might
drive a third species into extinction and so on,
producing an avalanche that in principle could span the whole
system. Because of the diverging mean avalanche size, the distribution
of extinction events would then be a power-law, similar to the
situation of 
thermodynamical systems at the point of a
phase-transition. Nevertheless, this mechanism, called
self-organized criticality, completely neglects external influences
that certainly are present.

On the contrary, as it has recently been shown,  a power-law
distribution of extinction events can appear even in a system in which
species are wiped out solely because of external
influences~\cite{Newman96b}. However, 
this effect depends crucially on influences that are imposed
on all species coherently.

From the point of view of a single species it does not really matter
whether it has to struggle with bad conditions imposed externally,
e.g., a global shift in temperature, or with bad conditions due to
heavy competition with other species. All that counts for a single
species is whether it can keep up with its environment or not.

A species goes extinct when its population decreases to zero. This can
happen for several reasons. One is a loss of habitat. Climatic or
tectonic changes affect the location and the size of a species'
habitat. If the size decreases rapidly, the species may not be able to
adapt fast enough to find a new niche. Then the population will drop
below a level at which it can sustain itself and the species will die
out. Another reason for species' extinction is the invasion of new
competitors or new 
predators. Competitors that invade a territory may be better adapted to
a niche than the species originally occupying this niche. In this
case, the population of the native species can be decimated
so effectively that it is wiped out. The same thing can happen because
of an invading predator superior to the defense mechanisms of
the species. Similarly, new parasites can significantly reduce the
population of a species and drive it to extinction.

From the species point of view, all the above cases can be subsumed
under the notion of stress. A species suffers stress of various kinds,
stress because of climatic changes, stress because of competition and
predation etc. If the stress exceeds the level a species can
sustain, it will go extinct.

\begin{figure}
\centerline{
        \epsfxsize=\columnwidth{\epsfbox{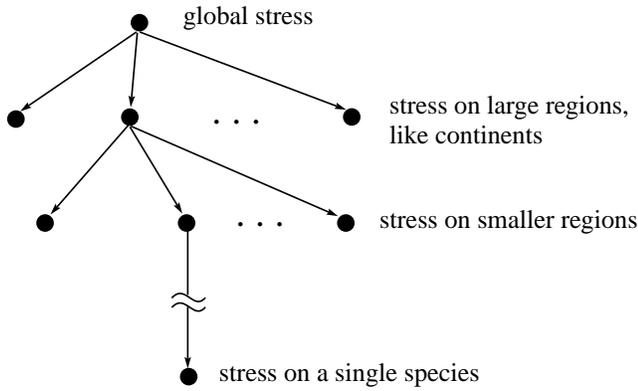}}
}

\caption{Stress is generated in a tree structure.}\label{Fig:hierarchy}  
\end{figure}

We are going to develop a model in which all causes for the extinction of
a species will be regarded as stress. Every species $i$ has a
threshold $x_i$, or in general 
a vector $\boldsymbol{x}_i$, against stress. If a species suffers a
stress $\eta_i> x_i$, or in the general case a vector of stresses
$\boldsymbol{\eta}_i$, where at least one component exceeds the
corresponding component of the threshold vector $\boldsymbol{x}_i$, it
dies out.
So far, this is a very general approach for a model in which species are
the smallest units considered, i.e.\ a model that does not work with
individuals or populations. 
Clearly, in such a model there will be stress on several scales. We
have global stress like a global shift in temperature due to a
slight change in the orbit of the Earth around the sun, or the impact
of a very large meteor. Then we have stress that spans large parts of
the Earth, 
e.g.\ a continent or a hemisphere, like the El Ni\~no phenomenon that
roughly spans the region about the tropical Pacific Ocean. And
finally, we have stress that affects smaller regions, or only
a single species. This leads us to a hierarchically ordered system of
environmental stresses. The simplest way to model it is to generate
stress in a tree structure, as it is shown in Fig.~\ref{Fig:hierarchy}.

On all scales, the stress can be abiotic or
biotic. This may sound a bit counter-intuitive, since abiotic changes
are usually taken as large-scale phenomena, and biotic factors are
usually taken as local phenomena. Abiotic changes happen often on a
global scale, like the above mentioned examples of the orbit shift of
the Earth or the meteor impact. But clearly there are more localized
events. A small meteor or a small vulcano may affect only a limited
number of species. If a species happens to live only in a very small
territory, and this territory gets destroyed by a meteor impact, the
species may be the only one that goes extinct because of the impact.

On the other hand, biotic phenomena are not necessarily 
localized. Although direct species
competition will usually be a local phenomenon, there can be
also global biotic phenomena. The composition of the atmosphere, for example,
depends strongly on biotic factors, and it can change significantly
due to biotic effects. 

So far, we have a model which represents the biosphere as a tree,
with species situated at the leafs, and
environmental stress generated at the nodes. Now we have to
choose the rules that determine how stress is generated and what
thresholds against stress the species are given. This is the crucial
part where we decide what mechanisms we want to investigate. If we
were interested mainly in coevolutionary effects, we would choose
rules that link the properties and actions of the species directly to
the generation of the stress. In such a model, for example, 
the global 
stress at time $t$ could be some sort of a sum over all
the adaptive moves of the species at time $t-1$. In this work,
however, we are mainly interested in the generic features we can
expect from the hierarchical structure of the biosphere. Therefore, we
will focus on a version of the model where the stresses and
the species' thresholds are simply random variables. Species'
interactions and abiotic effects can be so complicated 
and so unpredictable that in a first approximation we want to assume
them to be completely random. 

The model we study here is probably the simplest possible. Yet it has
some intriguing features which are very similar to characteristics
seen in the fossil record. To keep our model simple, we choose a
homogenous tree, with 
$l$ layers and $n$ subtrees per node. In general, of course, one has
to deal with inhomogenous trees. To each node of the final layer
we connect exactly one leaf, where we put $m$ species. 
An example of such a tree with $l=4$ and $n=2$ is displayed in
Fig.~\ref{Fig:subtrees}. The total number of stresses that have to be
generated in one time step is
\begin{equation}
  N_{\rm stress}=\sum_{i=0}^{l-1} n^i\,,
\end{equation}
and the total number of species in the model is
\begin{equation}
  N_{\rm species}=mn^{l-1}\,.
\end{equation}
Every species $i$ has a single threshold $x_i$, chosen at random from
the uniform distribution on the intervall $[0;1)$. At every node $j$,
the stress $\eta_j$ generated in one time step is a positive, real
random variable drawn from a distribution with probability
densitiy function (pdf) $p_j(x)$. It is a reasonable assumption to
expect smaller stresses to happen much more often than larger
stresses. Therefore, we use pdf's that fall off relatively fast with
$x\rightarrow \infty$. An Exponential or Gaussian decrease should be a
good choice, but the exact form of the pdf is not really
important. We choose the pdf's $p_j(x)$ at the beginning of the
simulation at random from some family of distribution functions and
keep this choice fixed throughout the course of the simulation.

Finally, we have to fix the way a species is affected by
stress generated on different levels of the tree. We simply take the
maximum of all the stress values generated at nodes that lie above the
species in the tree: if at any of these nodes a stress $\eta_j$ is
generated which exceeds the species threshold $x_i$, this species
goes extinct. It is then immediately
replaced by a new species with new random threshold. 

In addition to the extinction dynamic, we introduce some sort of
adaption. In agreement with our idea of a first, simple
model, the adaption is a random walk: in every time step, a
fraction $f$ of the species is selected at random and given new thresholds.

There are certainly some oversimplifications in this model, such as
the fixed number of species or the fact that all species have only one
trait. We will return to this later and explain why we can still
expect to cover the basic features of the extinction dynamic.

\section{Analysis}
The behaviour of the above introduced model can be understood to a
large extent from analytical calculations. But before we begin with our
analysis, we note that the mechanism for species extinction and
adaption presented here is similar to the one of the so-called
'coherent-noise' models 
introduced by Newman and Sneppen~\cite{Newman96a}. These models
display a distribution 
of extinction events that follows a power-law with exponent $\approx-2$,
which is in good agreement with the fossil record. For this reason, they have
already been used to study macroevolutionary
phenomena~\cite{Newman96b,Wilke97}. The difference
to our actual approach lies in the fact that we use a multitude of stresses
in a hierarchically ordered system, whereas in the coherent-noise
models there is only a single stress, acting on the whole system
at once. Therefore, in the previous works the idea of stress
imposed on the species has been linked to external influences like
meteor impacts and was opposed to coevolutionary effects.

Note that we have effectively a coherent-noise model at every leaf of the tree
if the number $m$ of species located at one leaf is large.

\subsection{The effective stress-distribution at a leaf of the tree}

Every leaf of the tree feels a stress-distribution which depends on
the distributions of the nodes above it. Let there be $N$ nodes above
a leaf. Then the $N$ stress values having influence on this leaf are
$N$ random variables $X_1, \dots, X_N$ with pdf's
$p_1(x), \dots, p_N(x)$. We have to calculate the pdf $p_{\max}(x)$ of the
random variable $X_{\max}=\max\{X_1, \dots, X_N\}$, i.e.,
\begin{equation}
  p_{\max}(x)\,dx=P(x\leq\max\{X_1, \dots, X_N\}<x+dx)\,.
\end{equation}
With the partition theorem we can write the probability on the
right-hand side as a weighted sum of conditional probabilities:
\begin{eqnarray}\label{eq:conditioning}
  &&P(x\leq\max\{X_1, \dots, X_N\}<x+dx)\nonumber \\
  &&\qquad= \sum_{i=1}^N P(x\leq\max\{X_1, \dots, X_N\}<x+dx\,\nonumber\\
  &&\qquad\qquad\qquad\qquad\qquad\quad\Big|\, x\leq X_i<x+dx)\nonumber\\
  &&\qquad\qquad    \times P(x\leq X_i<x+dx)\,.
\end{eqnarray}
The conditional probabilities read
\begin{eqnarray}\label{eq:condprob}
  &&P(x\leq\max\{X_1, \dots, X_N\}<x+dx\,\nonumber\\
  &&\qquad\qquad\qquad\qquad\qquad\quad\Big|\, x\leq X_i<x+dx)\nonumber\\
  &&\qquad=\frac{1}{P(x\leq X_i<x+dx)}\nonumber\\
  &&\qquad\quad\times P(x\leq\max\{X_1, \dots, X_N\}<x+dx\nonumber\\
  &&\qquad\qquad\qquad\qquad\qquad\wedge\, x\leq X_i<x+dx)\nonumber\\
  &&\qquad=\frac{ P(x\leq X_i<x+dx)\prod_{j=1, j\neq i}^N P(x > X_j)}
     {P(x\leq X_i<x+dx)} \nonumber\\
  &&\qquad=\!\!\prod_{j=1, j\neq i}^N\!\! P(x > X_j)\,.
\end{eqnarray}
After inserting Eq.~(\ref{eq:condprob}) into
Eq.~(\ref{eq:conditioning}) we find
\begin{eqnarray}
  &&P(x\leq\max\{X_1, \dots, X_N\}<x+dx)\nonumber \\
  &&\qquad= \sum_{i=1}^N P(x\leq X_i<x+dx)\!\!\!\prod_{j=1, j\neq i}^N 
  \!\!  P(x > X_j)\,.\nonumber\\
  &&
\end{eqnarray}
Consequently, for the pdf $p_{\max}(x)$ we have
\begin{eqnarray}\label{eq:pmax}
  p_{\max}(x)&&=\sum_{i=1}^N\, p_i(x)
   \!\!\! \prod_{j=1, j\neq i}^N\!\! P(x > X_j) \nonumber\\
    &&=\sum_{i=1}^N\, p_i(x)\!\!\! \prod_{j=1, j\neq i}^N
         \int\limits_0^x\! p_j(x')\,dx'\,.
\end{eqnarray} 
We are interested in the tail of $p_{\max}(x)$.
For coherent-noise models we know that a power-law distribution of
event-sizes will appear if 
the stress-distribution $p_{\rm stress}(x)$ satisfies
\begin{equation}
  \int\limits_\eta^\infty \!p_{\rm stress}(x)\,dx\approx Cp_{\rm
                stress}^\alpha(\eta)\quad\mbox{for $\eta\rightarrow\infty$}\,, 
\end{equation}
where $C$ and $\alpha$ are positive constants which depend on 
$p_{\rm stress}(x)$~\cite{Sneppen97}.
Therefore, we assume this condition to hold also for the
distributions $p_j(x)$ in Eq.~(\ref{eq:pmax}), with constants $C_j$
and $\alpha_j$, respectively. Then we can approximate the tail of
$p_{\max}(x)$ by
\begin{equation}
  p_{\max}(x)
    \approx\sum_{i=1}^N\, p_i(x)\!\!\! \prod_{j=1, j\neq i}^N
        \!\! \Big(1-C_j p_j^{\alpha_j}(x)\Big)\quad\mbox{for
    $x\rightarrow\infty$} \,. 
\end{equation}
We proceed further by taking only linear terms in $p_i(x)$ and obtain
\begin{equation}
  p_{\max}(x)
    \approx\sum_{i=1}^N\, p_i(x) \quad\mbox{for $x\rightarrow\infty$} \,. 
\end{equation}
For large $x$, this sum will be dominated by the $p_i(x)$ that is
falling off slowest. We say that
distribution $p_i(x)$ falls off slower than distribution $p_j(x)$ if
there exists a $x_0$ such that
\begin{equation}\label{eq:fall_off_slower}
  p_i(x)>p_j(x) \qquad \mbox{for all $x>x_0$.}
\end{equation}
For a set of reasonable stress-distributions it is always possible to identify
one that is falling off slowest according to this definition.

\begin{figure}

\centerline{
        \epsfxsize=\columnwidth{\epsfbox{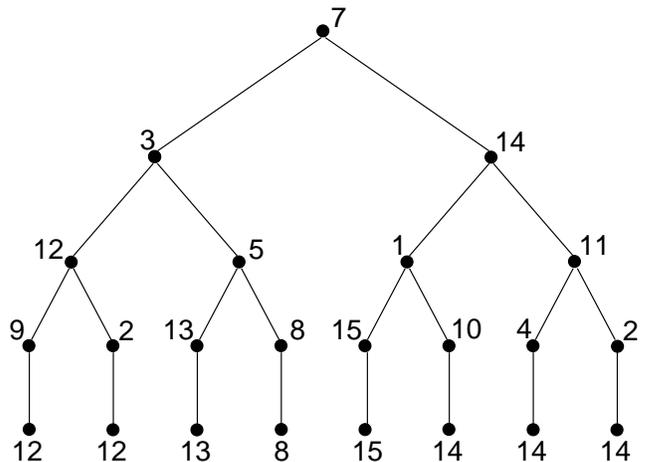}}
}

\caption{The tree breaks down into virtually independent parts in the
  limit of large stresses.} 
\label{Fig:subtrees}  
\end{figure}

The fact that the sum in Eq.~(\ref{eq:fall_off_slower}) will
asymptotically be dominated by a single term leads to the situation
depicted in Fig.~\ref{Fig:subtrees}. The tree breaks down into several
independent subsystems. The meaning of the numbers in
the figure will be explained in detail 
later. In a nutshell, they indicate how slow a stress distribution is
falling off. What interests us here is the breakup of the tree into
several independent parts in 
the regime of large stresses. If
these parts are not too small, 
they will behave like independent coherent-noise systems.

\subsection{An ensemble of a finite number of independent
  coherent-noise systems}

If the stress-distributions close to the root dominate the
behaviour of the system, the tree will break down into independent
coherent-noise systems, as we have mentioned above. Consequently, we
proceed with the calculation of the distribution of extinction events
in a system consisting of independent coherent-noise subsystems. In
the calculation, however, we will deviate slightly from the actual situation
in the tree model by assuming the subsystems to have each an
infinite size. This allows for an easy calculation, and the main
results should also hold for large but finite sizes.

In the case of an infinite system size, the event distribution of a
coherent-noise model possesses a power-law tail that extends to
arbitrary large events. Therefore, the task of calculating  the event
distribution of the compound system equals to the task of calculating
the sum of a finite number of nonidentically distributed random
variables with power-law tail. The latter can be treated
mathematically exact under relatively weak
assumptions~\cite{WilkeUnpublished}. But since the exact calculations
are too extensive to be included in this work, we will here give only
an intuitive argument about the tail behaviour of the sum. 

We begin with the sum of two positive, real random variables $X_1$ and $X_2$,
where the
pdf's $p_1(x)$ and $p_2(x)$ have a power-law tail $x^{-\tau_1}$ and
$x^{-\tau_2}$, respectively. We 
assume the pdf's to be continuous, non-singular, and reasonably
smooth. Under these conditions, we can write $p_1(x)$ and $p_2(x)$ in
the form
\begin{eqnarray}
  p_1(x)=\frac{f_1(x)}{(x+1)^{\tau_1}}\,,\\
  p_2(x)=\frac{f_2(x)}{(x+1)^{\tau_2}}\,,
\end{eqnarray}
where $f_1(x)$ and $f_2(x)$ are continuous, non-singular, and reasonably
smooth functions which tend towards a positive constant for
$x\rightarrow\infty$. The pdf $p_{\rm sum}(x)$ of the sum $X=X_1+X_2$ is the
convolution of $p_1(x)$ and $p_2(x)$:
\begin{eqnarray}
  p_{\rm sum}(x) &&= \int_0^x\! p_1(x')p_2(x-x')\,dx' \nonumber\\
 &&= \int_0^x\!\frac{f_1(x')}{(x'+1)^{\tau_1}}
          \frac{f_2(x-x')}{(x-x'+1)^{\tau_2}}\,dx'\,. \nonumber\\
 &&
\end{eqnarray}
After a change of the integration variable to $z=x'/x$ we obtain
\begin{eqnarray}
  p_{\rm sum}(x) &&=\int_0^1\!\frac{f_1(xz)}{(xz+1)^{\tau_1}}
          \frac{f_2(x(1-z))}{(x(1-z)+1)^{\tau_2}}x\,dx' \nonumber\\
  &&=x^{1-\tau_1-\tau_2}\int_0^1\!\frac{f_1(xz)}{(z+\frac{1}{x})^{\tau_1}}
          \frac{f_2(x(1-z))}{(1-z+\frac{1}{x})^{\tau_2}}\,dx'\,. \nonumber\\
  &&
\end{eqnarray}
For large $x$, there are two main contributions to this integral, at
$z\approx 0$ and at $z\approx 1$, which stem from the first and from
the second term in the denominator. Since the denominators will become
arbitrarily large for large $x$, we can assume the other terms to be
constant in the regions where the main contributions 
come from. Therefore, we find
\begin{equation}
  p_{\rm sum}(x) \approx x^{1-\tau_1-\tau_2} \Big[ C_1 x^{\tau_2-1} +
  C_2 x^{\tau_1-1}\Big]\,,
\end{equation}
where $C_1$ and $C_2$ are positive constants. Obviously for large $x$
the term with the largest exponent will dominate. Hence we have
\begin{equation}
  p_{\rm sum}(x) \sim x^{-\min\{\tau_1,\tau_2\}}\,.
\end{equation} 
This result can be easily extended to the case of an arbitrary finite
number of random variables with power-law tail by
iteration. Asymptotically, the tail of $p_{\rm sum}(s)$ will always be
dominated by the contribution from the term with the smallest
exponent.

Back to the ensemble of infinitely large coherent-noise systems, we
find that it will 
display power-law distributed event sizes, as its 
single constituents do. If the subsystems' stress-distributions are
functionally different, the exponent of the compound system's event
distribution will be the smallest of the subsystems' exponents.

The above result should also hold in the situation of finite
coherent noise systems, as long as their total number is small
compared to their typical size.

\section{Trees with random stress distributions}

We argue above that in the limit of large
stresses the tree will break down into subsystems, virtually
independent of each other. The behaviour of our model depends
heavily on the size of the parts we find. If the different parts are
all very small, the system will loose its coherent-noise
characteristics. Instead of a power-law distribution the
extinction events will then follow a gaussian distribution because of
the central-limit theorem. Therefore, in this section we will study
the distribution of the subsystems' sizes that arises if we randomly assign
stress distributions to the tree's nodes.

We assume that the propability for a certain stress distribution to be
assigned to a certain node does not depend on the position of the node
in the tree. In other words, we use the same set of
stress distributions on all levels of the tree. Furthermore, we assume
that for any two stress distributions we use we can identify one of
the two that falls off faster than the other one.
Under these conditions, we can study the structure of such trees by
simply assigning integers to the nodes of the tree, where larger
integers stand for distributions that are falling off slower. If the set of
possible stress distributions is infinite, the probability of finding
two nodes with the same distribution is zero. Consequently, in a tree
with $n$ nodes, we will assign every integer from $1\dots n$ to exactly
one node. This is displayed in Fig.~\ref{Fig:subtrees} for a tree with
15 nodes. For every leaf $i$
of the tree we can then define a characteristic number $a_i$. This
number is the maximum of the nodes' numbers encountered on the way
from the leaf up 
to the root. All the leafs with the same characteristic
number belong to the same subsystem. In the example of
Fig.~\ref{Fig:subtrees}, we have five subsystems in total. Three of them 
contain only one leaf, one contains two and one contains three leafs.

\begin{figure}
\centerline{
        \epsfxsize=\columnwidth{\epsfbox{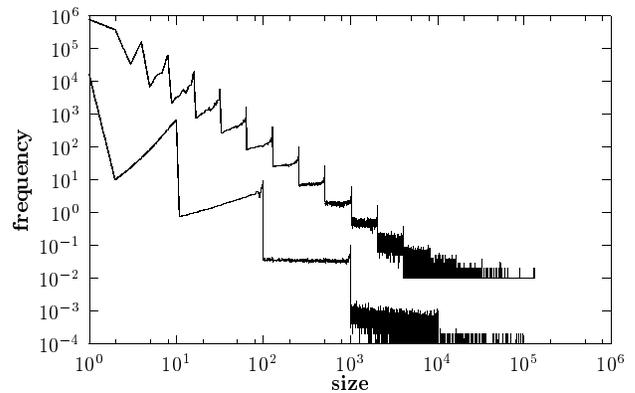}}
}

\caption{The expected frequency for the occurence of large independent
  subsystems decreases as a sawtooth function that follows
  approximately a power-law with 
  exponent~$-2$. The upper curve stems from a tree with $l=18$ and
  $n=2$. It has been rescaled by a factor of 100 so as
  not to overlap with the lower curve. The lower curve stems from a tree
  with $l=6$ and $n=10$.} 
\label{Fig:parts}  
\end{figure}

In general, we are interested in the distribution of subsystems
arising in 
large trees. Therefore, we have done simulations in which we have
several thousand times assigned
random integers to the nodes of a large tree. For every single
realization of the tree, we have computed a 
histogram of the frequency of the different parts' sizes. Finally, we
have calculated the average of all the histograms.
Fig.~\ref{Fig:parts} shows the result of such simulations for two
different trees with 10000 histograms each. We find the
expected frequency $f(k)$ of large independent 
parts in the tree decreasing as a sawtooth function that follows
approximately a power-law with 
exponent~$-2$, independent of $l$ and $n$.  The sharp peaks in the
distribution 
arise whenever the size of a complete subtree is reached. Therefore, we
observe in Fig.~\ref{Fig:parts}, e.g., the peaks in the distribution of
the tree with $n=10$ appearing at powers of 10.

The
power-law can be explained easily if we assume the main
contributions to come from complete subtrees. The expected frequency 
$f(k)$ to find an independent subtree with $b$ layers, which
corresponds to a subsystem of size $k=n^b$, can be written as the
number of such subtrees in the whole system, $N(b)$, times the probability that
any of these subtrees will be independent of the rest, $P(b)$. Hence
we write
\begin{equation}
  f(n^b)=N(b)P(b)\,.
\end{equation}
The number of subtrees of size $n^b$ is $N(b)=n^{l-b}$. For the
probability $P(b)$ we find
\begin{equation}
  P(b)=\Big(l-b+\sum_{i=0}^{b-1}n^i\Big)^{-1}\,,
\end{equation}
which is simply the probability for the integer assigned to the node
at the root of the 
subtree to be larger than all the other integers which are assigned to
the remaining 
nodes of the subtree and to the nodes above the subtree. 
If we increase $b$ by one, we get $N(b+1)=n^{l-b-1}=N(b)/n$. With
slightly more effort, we find also
\begin{eqnarray}
  P(b+1)&=&\Big(l-b-1+\sum_{i=0}^{b}n^i\Big)^{-1}  \nonumber\\
 &=&\Big(l-b+n\sum_{i=0}^{b-1}n^i\Big)^{-1}
 \approx \frac{1}{n}P(b)\,.
\end{eqnarray}
Therefore, we can write
\begin{equation}
 f(nk)\approx \frac{N(k)}{n}\frac{P(k)}{n} =n^{-2}f(k)\,,
\end{equation}
which implies $f(k)\sim k^{-2}$.

The peaks in Fig.~\ref{Fig:parts} appear whenever the size of a
complete subtree is reached, as we have noted above. This means they
are connected to the extremely regular structure of the trees we use
in this work. Therefore, we are currently investigating trees with irregular
structure. For these trees, the spikes disappear and, in log-log plot,
the function $f(k)$ becomes almost a straight line with slope -2. From
the simulations we have done so far, we can say that this result is
very general and seems to be independent of the special trees' properties.

\section {Simulation results}

Since we are interested in the typical behaviour of our model, we have
to do many simulation runs with different tree sizes and different
stress distributions at the tree's nodes. But the simulation of
large trees is very slow, and therefore it is hard to get a good
sample of the parameter-space. To overcome this difficulty we have
also done simulations based on the arguments of
the previous sections. As we have seen there, in the limit of large
stresses it is possible 
to map the leafs of the tree onto a system consisting of several
independent coherent-noise models, with the sizes $k$ of these subsystems
distributed according to $k^{-2}$.

\begin{figure}

\centerline{
        \epsfxsize=\columnwidth{\epsfbox{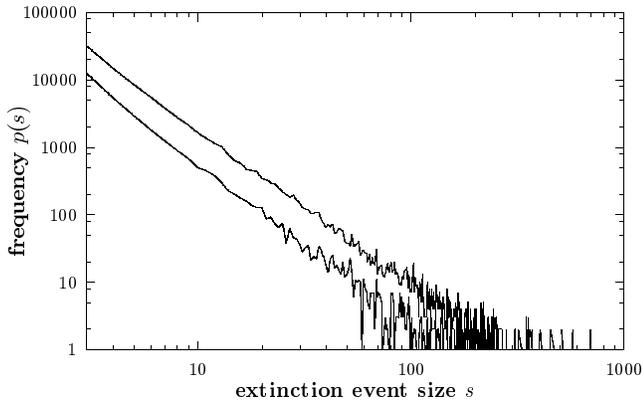}}
}

\caption{The frequency of extinction events in the tree model and in
        an ensemble of coherent-noise models. The lower curve stems
        from the simulation of a tree with 
        $l=5$, $n=10$ and $m=1$, which amounts to a total of $10^5$
        species. Stress 
        distributions were assigned at random to the nodes of the
        tree. We used exponentially distributed stress with $\sigma$
        between 0.03 and 0.05. The upper curve corresponds to the
        simulation of an ensemble of coherent-noise models with a total
        of $10^5$ species, and with the sizes $k$ of the
        subsystems distributed according to $k^{-2}$. As
        stresses we used the maximum of 5 exponentially distributed
        random variables.} 
\label{Fig:comparison}  
\end{figure}

In Fig.~\ref{Fig:comparison} we show a comparison between the full
simulation and the approximation. To come as close as possible to the
full simulation,  we use the maximum of 5 independent,
exponentially distributed random variables as stresses for the
independent coherent-noise models, since for the tree we have likewise chosen
$l=5$ and exponential stress-distributions.  Clearly the behaviour of the
approximation is close to the one of the full simulation, which
verifies the analytical reasoning of the previous sections. Both
simulations display power-law distributed extinction events. For the
full tree, we find an exponent $\tau_{\rm tree}=2.35\pm0.05$, while
for the approximation, we find $\tau_{\rm approx}=2.30\pm0.05$. If we
consider the high level of abstraction from the tree to an ensemble
of coherent-noise systems, this agreement is excellent.

Note that in comparison to a normal coherent-noise model with only a
single stress variable, the tree model produces a significantly larger
exponent $\tau$ (If we run a normal coherent-noise model with the
stress distribution of the 
approximaton in Fig.~\ref{Fig:comparison}, we get an exponent
$\tau\approx 1.8$). 
The increased exponent $\tau$ has its origin in the distribution of
the subsystems' sizes. The sizes scale themselves, thus modifying the
scale-invariant behavior of the ensemble, compared to the one of a
single coherent-noise system. 

\section {Discussion}
We have presented a model of large-scale evolution and extinction that
combines biotic and abiotic causes for extinction within a single
mathematical framework. Furthermore, the model takes into account the
hierarchical structure of the biosphere. To the best of our knowledge,
the implications of environmental changes happening on different
scales have not been studied previously in macroevolutionary
models. Despite the 
choice of completely random environmental changes, the model has some
interesting features. The distribution of
extinction events follows a power-law with exponent in the region of
2 (note that the exponent depends on the choice of the stress
distribution, as it is the case with coherent-noise models). From the
fossil record, a power-law distribution 
with exponent $\tau\approx 2$ is reported for the extinction event
sizes of taxonomical families 
\cite{Sole96,Newman96b}. Moreover, it is interesting to observe the
breakup of the tree into subsystems with sizes $k$ distributed
according to $k^{-2}$. The power-law distribution of the subsytem
sizes implies that even in very large trees we will find large
subsystems, governed mainly by only a single stress
distribution. Intuitively, we
would expect the subsystems to have roughly similar sizes, and to
enter the dynamic of the whole system on an equal basis. But we
observe exactly the opposite. The subsystems' sizes are
scale-invariant, thus producing a scale-invariant distribution of
contributions to the overall system's behavior. In particular, only a
small number of large subsystems produces events on large scales.
This might be an
explanation for the fact that in such 
large and complex systems like the biosphere we find usually smooth
frequency distributions of typical objects or events. 

The model we have studied in this work is certainly oversimplified. For
that reason, we will close this paper with some remarks about 
extensions to the model that should be examined in a next step closer
to biological reality. First of all, it is certainly a severe restriction 
to keep the number of species fixed throughout the
simulation. Nevertheless, this is a restriction used very
often in models of macroevolution~\cite{Peliti97}. Only recently, work
has been done where a change in biodiversity is
considered~\cite{Head97,Wilke97}. The behaviour of the model we study here is
governed by the coherent-noise dynamic. For this dynamic, it has
been shown that it can be generalized to include a variable system size
without loss of it's main features~\cite{Wilke97}. Therefore, we
believe a fixed system-size can be justified in the present work. 
It should be possible to extend our tree model to a model with variable
system size. Another severe restriction is the usage of only one
trait. But here a similar argument holds as in the case of the fixed
number of species. A multi-trait version of the original
coherent-noise model has already been studied~\cite{Newman97b}. It
behaves very similar to the single-trait version.

Finally, we want to discuss the way we compute the stress on a single
species out of the multitude of stress values, generated at the
different levels of the tree. Throughout this paper, we have used the
maximum of the stress values. This allows for an easy and very general
analytical investigation. Another natural choice, however, would be to
sum up all the stresses. We have also done some simulations in this
fashion. The behavior of the system remains roughly the same. This
happens because in a finite sum of non-identically distributed random
variables, we expect large values to be dominated by a single term of
the sum, similar to the case of the maximum of several random
variables. For the sum of exponentially distributed random variables, an easy
calculation shows that this conjecture is indeed true. With some more effort,
we can prove the same for the sum of power-law
distributed random variables, as we have already done in this
paper. Nevertheless, in the general case with arbitrary distributions,
the conjecture is hard to demonstrate.

\end{document}